%% file: dark2004.tex
\documentclass[multphys,vecphys]{svmult}

\newcommand {\apgt} {\ {\raise-.5ex\hbox{$\buildrel>\over\sim$}}\ }
\newcommand {\aplt} {\ {\raise-.5ex\hbox{$\buildrel<\over\sim$}}\ }

\usepackage{makeidx}         % allows index generation
\usepackage{graphicx}        % standard LaTeX graphics tool
\usepackage{multicol}        % used for the two-column index
\usepackage[bottom]{footmisc}% places footnotes at page bottom
\makeindex             % used for the subject index
\begin{document}

\title*{How Dark is `Dark'? Electromagnetic Interactions in the Dark Sector}
\author{Kris Sigurdson}
\institute{California Institute of Technology, Mail Code 130-33, Pasadena, CA 91125 USA
\texttt{ksigurds@tapir.caltech.edu}}
%\and Name and Address of your Institute \texttt{name@email.address}}
\maketitle
\begin{abstract}
	We review the physical and cosmological consequences of two possible electromagnetic couplings to the dark sector: ({\rm i}) a neutral lightest dark-matter particle (LDP) with nonzero electric and/or magnetic dipole moments and (ii) a charged next-to-lightest dark-matter particle (NLDP) which decays to a neutral LDP.  For scenario (i) we find that a relatively light particle with mass between a few MeV and a few GeV and an electric or magnetic dipole as large as $\sim 3\times 10^{-16}e$~cm (roughly $1.6\times10^{-5}\,\mu_B$) satisfies experimental and observational bounds.  In scenario (ii), we show that charged-particles decaying in the early Universe result in a suppression of the small-scale matter power spectrum on scales that enter the horizon prior to decay. This leads to either a cutoff in the matter power spectrum, or if the charged fraction is less than unity, an effect in the power spectrum that might resemble a running (scale-dependent) spectral index in small-scale data. 
\end{abstract}

%%%%%%%%%%%%%%%%%%%%% Section 1 %%%%%%%%%%%%%%%%%%%

\section{Motivation}
%%%%%%%%%%%%%%%%%%%%% Section 1 %%%%%%%%%%%%%%%%%%%

	The origin of the missing `dark' matter in galaxies and clusters of galaxies has been an outstanding problem for over 70 years, since Zwicky's measurement of the masses of extragalactic systems 
\cite{Zwicky:1933}. 
	Recent cosmological observations not only tell us how much dark matter exists but also that it must be nonbaryonic 
\cite{CMBconstraints} --- 
	it is not one of the familiar elementary particles contained within the standard model of particle physics.  Dark matter is a known unknown.  We do not know what the underlying theory of dark matter is, what the detailed particle properties of it are, nor the particle spectrum of the dark sector. 

	Promising candidates for the lightest dark-matter particle (LDP) ---those that appear in minimal extensions of the standard model and are expected to have the required cosmological relic abundance --- are a weakly-interacting
massive particle (WIMP), such as the neutralino, the lightest mass eigenstate from the superposition of the supersymmetric 
partners of the $U(1)$ and $SU(2)$ neutral gauge bosons and of the 
neutral Higgs bosons~
\cite{Jungman:1995dflarsrev}, 
	or the axion
\cite{axionreviews}. 
	There is a significant theoretical literature on the properties 
	and phenomenology of these particles, and there are ongoing 
	experimental efforts to detect these particles.  

	There has also been a substantial phenomenological effort toward 
	placing model-independent limits on the possible interactions 
	of the LDP.  For instance, significant constraints have been 
	made to dark-matter models with strong interactions 
\cite{Starkman:1990nj} 
	and self-interactions 
\cite{Carlson:1992Spergel:1999mh},
        and various models with unstable particles have been investigated 
\cite{DecayingDM}.  
	Electromagnetic interactions have also been considered, 
	and models with stable charged dark matter have been ruled out 
\cite{Gould:1989gw} 
	while there are strong constraints on millicharged dark-matter 
	models 
\cite{Davidson:2000hfDubovsky:2003yn}.

	Dark matter is so called because the coupling of it to photons 
	is assumed to be nonexistent or very weak.  
	Here we ask the question, ``How dark is `dark'?'', 
	and review several recent investigations that consider 
	the physical and cosmological constraints to and effects 
	of the electromagnetic interactions of the LDP and 
	the next-to-lightest dark-matter particle (NLDP).  
	In particular, in Section~
\ref{sec:ddm} 
	we discuss the consequences of a neutral LDP with nonzero 
	electric and/or magnetic dipole moments 
\cite{Sigurdson:2004zp}, 
	and in Section~
\ref{sec:chargeddecay} 
	we discuss the cosmological effects of a charged NLDP 
	which decays to a neutral LDP 
\cite{Sigurdson:2003vy,Profumo:2004qt}.

%%%%%%%%%%%%%%%%%%%%%%%% Section 2 %%%%%%%%%%%%%%%%%%%%
\section{Dark-Matter: Electric and Magnetic Dipole Moments}
%%%%%%%%%%%%%%%%%%%%%%%% Section 2 %%%%%%%%%%%%%%%%%%%%

%%%%%%%%%%%%%%%%%%%%% Fig.1 %%%%%%%%%%%%%%%%%%%%
\label{sec:ddm}
\begin{figure}[h]
\begin{center}
\resizebox{11.65cm}{!}{\includegraphics{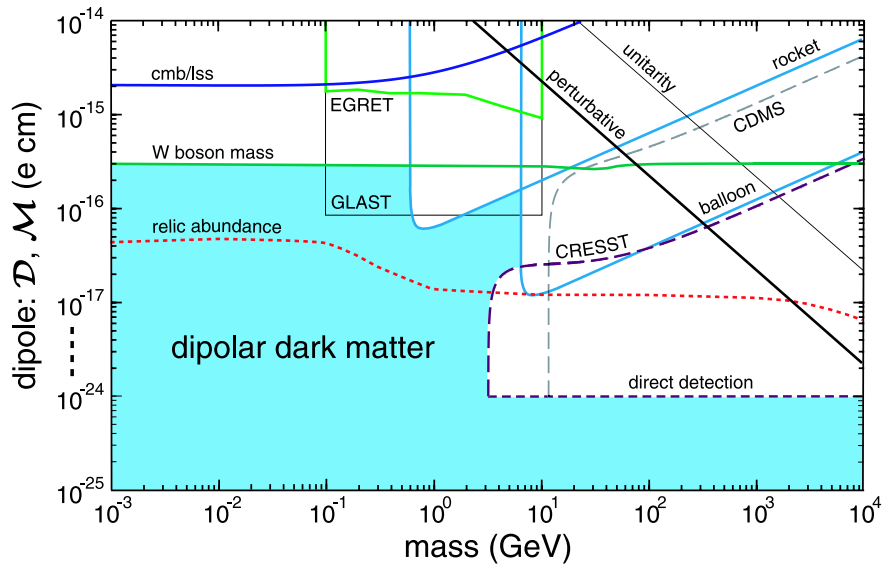}}
\caption{\label{fig:concord} The constraints to 
	$[m_\chi,\,({\cal D,\,M})]$ that come from present-day searches
	and experiments.  Viable candidates must lie in the shaded region.
 	The short-dashed ``relic abundance'' curve shows where the dark matter 
	would have a cosmological relic abundance 
	$\Omega_\chi h^2=0.135$, assuming standard freezeout 
	of annihilations via the dipole coupling to 
	$\gamma$ and no $\chi$-$\bar{\chi}$ asymmetry.
}
\end{center}
\end{figure}
%%%%%%%%%%%%%%%%%%%%% Fig.1 %%%%%%%%%%%%%%%%%%%%

	In this section we consider the possibility that 
	the dark matter possesses an electric or magnetic dipole moment. 
	The result of Ref.~
\cite{Sigurdson:2004zp}, 
	illustrated in
Fig.~\ref{fig:concord}, 
	is that a Dirac particle with an electric or magnetic
dipole moment of order $\sim 10^{-17} e~$cm with a mass between an MeV and a
few GeV can provide the dark matter while satisfying all experimental and
observational constraints.\footnote{We quote numbers for both
the electric and magnetic dipole moments in units of $e$~cm, where $e$ is the
electron charge.  For reference, the Bohr magneton $\mu_B = e\hbar/2
m_e=1.93\times10^{-11}\,e$~cm in these units.} 

	The effective Lagrangian for coupling of a Dirac 
	fermion $\chi$ with a magnetic dipole moment ${\cal M}$ 
	and an electric dipole moment ${\cal D}$ to the
	electromagnetic field $F^{\mu\nu}$ is
\begin{eqnarray}
     {\cal L}_{\gamma\chi} = 
          -\frac{i}{2}\bar{\chi}\sigma_{\mu\nu}({\cal M} 
          + \gamma_{5}{\cal D})\chi F^{\mu\nu}.
\label{Lint}
\end{eqnarray}
	Below we summarize various physical and cosmological limits 
	to the form of interaction shown in Eq.~(\ref{Lint}).  
	For further details of these limits see Ref.~
\cite{Sigurdson:2004zp}.

%%%%%%%%%%%%%%%%%%%%% Section 2.1 %%%%%%%%%%%%%%%%%%%%%%
\subsection{Dark Matter: Annihilation and Relic Abundance}
\label{section3}
%%%%%%%%%%%%%%%%%%%%% Section 2.1 %%%%%%%%%%%%%%%%

\begin{figure}[h]
\begin{center}
$\begin{array}{c@{\hspace{1in}}c}
\resizebox{3.3cm}{!}{\includegraphics{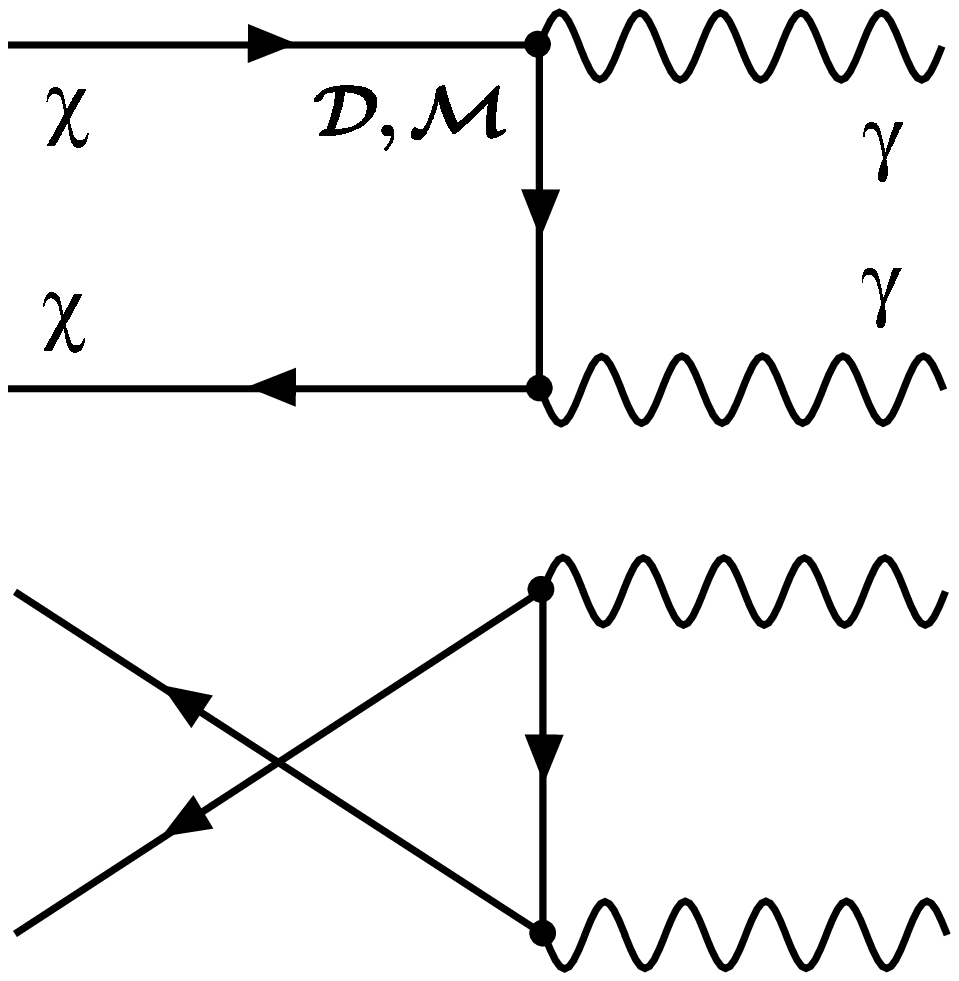}} &
\raisebox{0.6cm}{\resizebox{3.3cm}{!}{\includegraphics{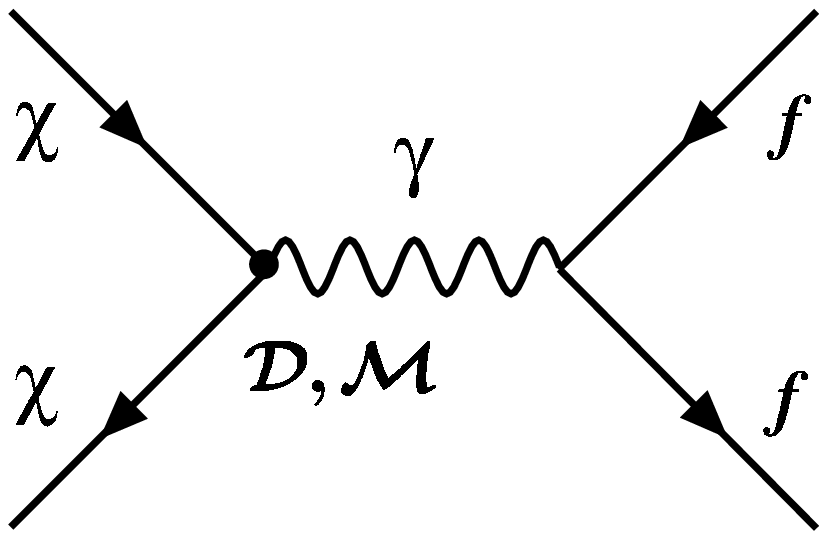}}} \\ [0.05cm]
\mbox{\bf (a)} & \mbox{\bf (b)}
\end{array}$
\end{center}
\caption{Feynman diagrams for (a) annihilation of a $\chi$--$\bar{\chi}$
	pair to two photons and (b) $\chi$--$\bar{\chi}$ annihilation 
	to charged $f$--$\bar{f}$ pairs.}
\label{fig:ddannhil}
\end{figure}

	We assume $\chi$ particles exist in thermal equilibrium 
	in the early Universe  and their dipole interactions freeze 
	out when $T$ drops below $m_\chi$.  
	Their cosmological relic abundance is  
$\Omega_\chi h^2 \simeq 3.8\times10^{7}\left({m_\chi/m_p}\right){\ln\left(A/\sqrt{\ln  A}\right) / A}$ where $A = 0.038 \sqrt{g_*}m_{pl} m_\chi (\sigma_{\rm ann} v)$ (see, e.g., Eq. (5.47) in Ref. 
\cite{KolbTurnerbook}).
	Here, $g_*$ is the effective number of relativistic degrees of freedom
at the freezeout temperature $T_f \sim m_\chi/A$. $\chi$--$\bar{\chi}$ pairs annihilate to either photons or charged pairs through the diagrams shown in Fig.
\ref{fig:ddannhil} and $\sigma_{\rm ann}v=\sigma_{\chi\bar\chi \to 2\gamma} v + \sigma_{\chi\bar\chi \to f\bar f} = ({\cal D}^4 + {\cal
     M}^4) m_\chi^2 / 2\pi + N_{\rm eff} \alpha ({\cal
     D}^2 + {\cal M}^2)$, where $N_{\rm eff}$ is the number of $f$--$\bar{f}$ pairs with $m_f<m_\chi$.  If  $\Omega_\chi h^2 = 0.135$ then 
$({\cal D}^2 +{\cal M}^2)^{1/2} \simeq 1.0 \times 10^{-17}\, e$~cm for  $m_\chi
\sim 1~$GeV, as shown in
Fig.~
\ref{fig:concord}.  
	The present-day mass density of $\chi$ particles might differ from these estimates if other interactions are significant or there is a $\chi$--$\bar{\chi}$ asymmetry.

%%%%%%%%%%%%%%%%%%%%% Fig.2 %%%%%%%%%%%%%%%%%%%%
%%%%%%%%%%%%%%%%%%%%% Fig.2 %%%%%%%%%%%%%%%%%%%%

%%%%%%%%%%%%%%%%%%%%% Section 2.2 %%%%%%%%%%%%%%%%%%%%%%
\subsection{Direct Detection}
\label{section4}
%%%%%%%%%%%%%%%%%%%%% Section 2.2 %%%%%%%%%%%%%%%%%%%%%%

\begin{figure}[h]
\begin{center}
$\begin{array}{c@{\hspace{1in}}c}
\raisebox{0.5cm}{\resizebox{3.3cm}{!}{\includegraphics{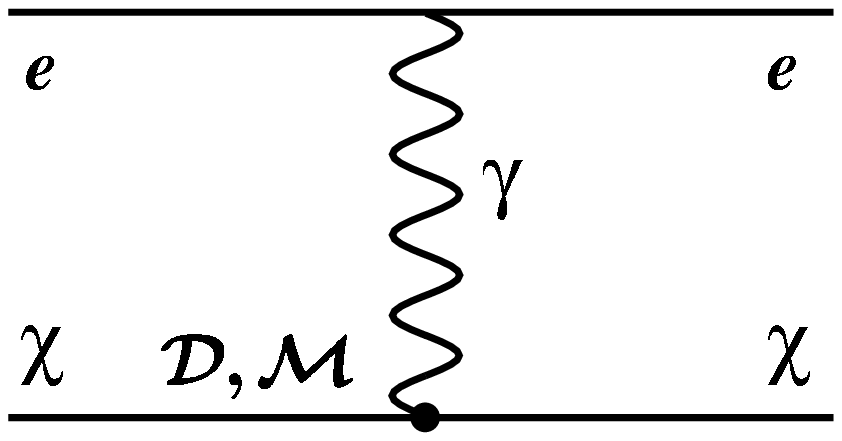}}} &
\resizebox{4.2cm}{!}{\includegraphics{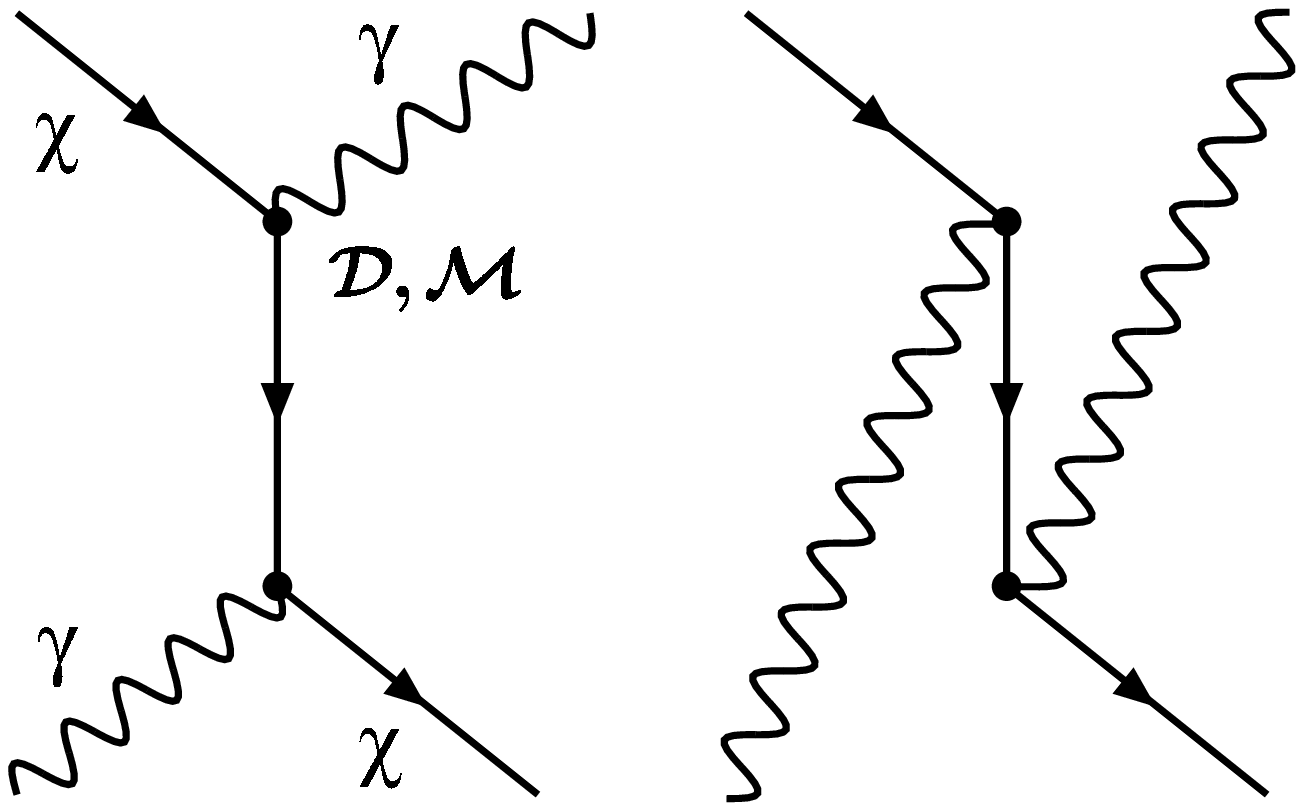}} \\ [0.05cm]
\mbox{\bf (a)} & \mbox{\bf (b)}
\end{array}$
\end{center}
\caption{Feynman diagrams for (a) the scattering of $\chi$ particles 
	by charged particles and (b) $\gamma$-$\chi$ scattering.}
\label{fig:dn}
\end{figure}

	In the nonrelativistic limit, the differential cross section 
	for the process shown in Fig.~\ref{fig:dn}a is  
	${d \sigma}/{d \Omega} = {Z^2 e^2 \left({\cal D}^2+{\cal M}^2\right)}/[{8\pi^2 v^2 (1 - \cos\theta)]}$, 
	where $v$ is the relative velocity.  Roughly speaking, $\sigma\sim (Ze)^2 ({\cal D}^2+{\cal M}^2)/2\pi v^2 \simeq 6.4 \times
10^{-32}\,Z^2 ({\cal D}_{17}^2+{\cal M}_{17}^2)$~cm$^2$, using
$v\sim10^{-3}\,c$.\footnote{$[{\cal D}_{17},\,{\cal M}_{17}]=[{\cal D,\, M}]/(10^{-17}~e~{\rm cm})$} Current null searches  in germanium detectors \cite{CDMS04} thus require $({\cal D}_{17}^2+{\cal M}_{17}^2)^{1/2}\aplt 10^{-7}$ at $m_\chi\sim10$~GeV --- improving upon previous limits \cite{Pospelov:2000bq}.

%%%%%%%%%%%%%%%%%%%%% Fig.3 %%%%%%%%%%%%%%%%%%%%

%%%%%%%%%%%%%%%%%%%%% Fig.3 %%%%%%%%%%%%%%%%%%%%

	However, for large enough dipole moments, $\chi$ particles 
	will lose energy in the rock/shielding above the detector 
	and evade detection in underground experiments. 
	Given a shielding thickness $L$ (in meters water equivalent), 
	we obtain the lower bound  
${\cal D}^2+{\cal M}^2 > [\frac{1}{2} m_\chi v^2 - 
\frac{1}{4} \frac{m_\chi m_d}{\mu[m_\chi,m_d]^2} E_{\rm th}]\times [{e^2 \over 2 \pi} L \sum_i f_i Z_i^2 
{\mu[m_\chi, m_i]^2 \over  m_i^2} ( 1 + \sqrt{\frac{m_i}{2 m_\chi}})]^{-1}$, where $\mu[m_\chi,m]=m_\chi
m(m_\chi+m)^{-1}$ is the reduced mass, $m_d$ is the mass of detector nuclei, $E_{\rm th}$ is the threshold nuclear-recoil energy, the index $i$ sums over the composition of the shielding material, and
$f_i$ is the fractional composition by weight.  The most restrictive constraints for large dipoles actually come from shallow experiments with null results such as the Stanford
Underground Facility run of the Cryogenic Dark Matter Search 
\cite{Akerib:2003px}, and the Cryogenic Rare Event Search with Superconducting Thermometers 
\cite{cresst}. Airborne experiments, in particular the balloon experiment of Ref.~\cite{Rich:1987st} and the rocket experiment of Ref.~\cite{McCammon:2002gb} provide important complimentary constraints.   The bounds due to all these experiments are shown in Fig.~\ref{fig:concord}.

%%%%%%%%%%%%%%%%%%%%% Section 2.3 %%%%%%%%%%%%%%%%%%%%%%
\subsection{Constraints from Precision Measurements}
\label{section5}
%%%%%%%%%%%%%%%%%%%%% Section 2.3 %%%%%%%%%%%%%%%%%%%%%%

\begin{figure}[ht]
\begin{center}
\resizebox{4.95cm}{!}{\includegraphics{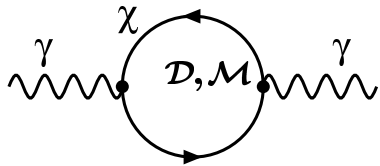}}
\caption{One-loop correction to the photon self-energy induced 
	by dipole moments $\cal{M},\cal{D}$ of the dark-matter 
	particle.}
\label{fig:photon-prop}
\end{center}
\end{figure}

	In Ref.~
\cite{Sigurdson:2004zp} 
	the effect of the dipole interaction on the anomalous magnetic 
	moment of the muon, standard model EDMs, and corrections to Z-pole 
	observables were considered. The strongest constraint was found 
	to arise from the contribution of $\chi$ particles to the running 
	of $\alpha$.  Such running affects the relationship 
	between $G_F$, $m_{W}$, and the value of $\alpha$ 
	at zero momentum: 
$m_W^2 = (\pi \alpha)/(\sqrt{2} G_F)\left[(1 - m_W^2/m_Z^2)(1-\Delta r)\right]^{-1}$.  
	In the standard model $\Delta r^{SM}=0.0355\pm 0.0019\pm 0.0002$ 
	while experimentally $\Delta r^{Exp}=0.0326\pm 0.0023$ 
	yielding $\Delta r^{New}<0.003$ with 95\% confidence.  
	The dipole interaction contributes to $\Delta r$ via
	the diagram in Fig.~
\ref{fig:photon-prop} 
	and so $({\cal D}^2+{\cal M}^2)^{1/2} \aplt 3\times 10^{-16}~e$~cm 
	is required, as shown in
Fig.~\ref{fig:concord}.  

%%%%%%%%%%%%%%%%%%%%% Fig.4 %%%%%%%%%%%%%%%%%%%%
%%%%%%%%%%%%%%%%%%%%% Fig.4 %%%%%%%%%%%%%%%%%%%%

%%%%%%%%%%%%%%%%%%%%% SubSection 2.4 %%%%%%%%%%%%%%%%%%%%%%
\subsection{Direct Production}
%%%%%%%%%%%%%%%%%%%%% Section 2.4 %%%%%%%%%%%%%%%%%%%%%%

	Missing-energy searches for light ($m_\chi\aplt 1$ GeV) 
	dark matter in rare $B^+$ decays was suggested in Ref.~
\cite{Bird:2004ts} 
	where $Br(B^+\to K^+ + {\rm invisible})\aplt 10^{-4}$  was derived.  
	This limit requires ${\cal D}\aplt 3.8\times 10^{-14}~e$~cm 
	for $m_\chi<(m_{B^+}-m_{K^+})/2=2.38$ GeV.  
	Similarly, rare $K^+$ decays lead to the limit 
	${\cal D}\aplt 1.5\times 10^{-15}~e$~cm for  
	$m_\chi<(m_{K^+}-m_{\pi^+})/2=0.18$ GeV. 
	These constraints are not yet competitive with other constraints 
	shown in Fig.~
\ref{fig:concord}.

	In order to limit dipole couplings using collider experiments 
	an expression for the rate 
	$f\bar f\to X \bar\chi \chi$, where $X$ is some set of visible 
	final-state particles, is necessary. Naive application 
	of the effective Lagrangian in Eq.
(\ref{Lint}) is invalid because perturbation
	theory breaks down when the energy scale of a process satisfies 
	${\cal E} \apgt 1/{\cal D}$. 
	Missing-energy searches from L3 (${\cal E} \approx 200$ GeV) 
	and CDF (${\cal E}=1.8$ TeV) can not be directly applied 
	to effective dipole moments ${\cal D}>10^{-16}~e$~cm and 
	${\cal D}>10^{-17}~e$~cm,
	respectively, unless a high-energy theory is specified.

%%%%%%%%%%%%%%%%%%%%% SubSection 2.5 %%%%%%%%%%%%%%%%%%%%%%
\subsection{Constraints from Large-Scale Structure and the CMB}
\label{section6}
%%%%%%%%%%%%%%%%%%%%% SubSection 2.5 %%%%%%%%%%%%%%%%%%%%%%

	A dipole moment induces a coupling of
	the dark matter to the primordial plasma by scattering via
	the diagrams shown in Fig.~
\ref{fig:dn}.  Dark matter couples
	to the plasma at early times, and subsequently decouples.  
	When coupled to the plasma, short-wavelength modes 
	of the dark-matter density field will grow less quickly relative 
	to the standard case.
	The long-wavelength modes that enter the horizon after dark matter 
	decoupling remain unaffected.  
	The full calculation in cosmological perturbation theory 
	is provided in Ref.~
\cite{Sigurdson:2004zp}.

%%%%%%%%%%%%%%%%%%%%% Fig.5 %%%%%%%%%%%%%%%%%%%%
\begin{figure}[htb]
\begin{center}
$\begin{array}{c@{\hspace{0.6in}}c}
\resizebox{7.73cm}{!}{\includegraphics{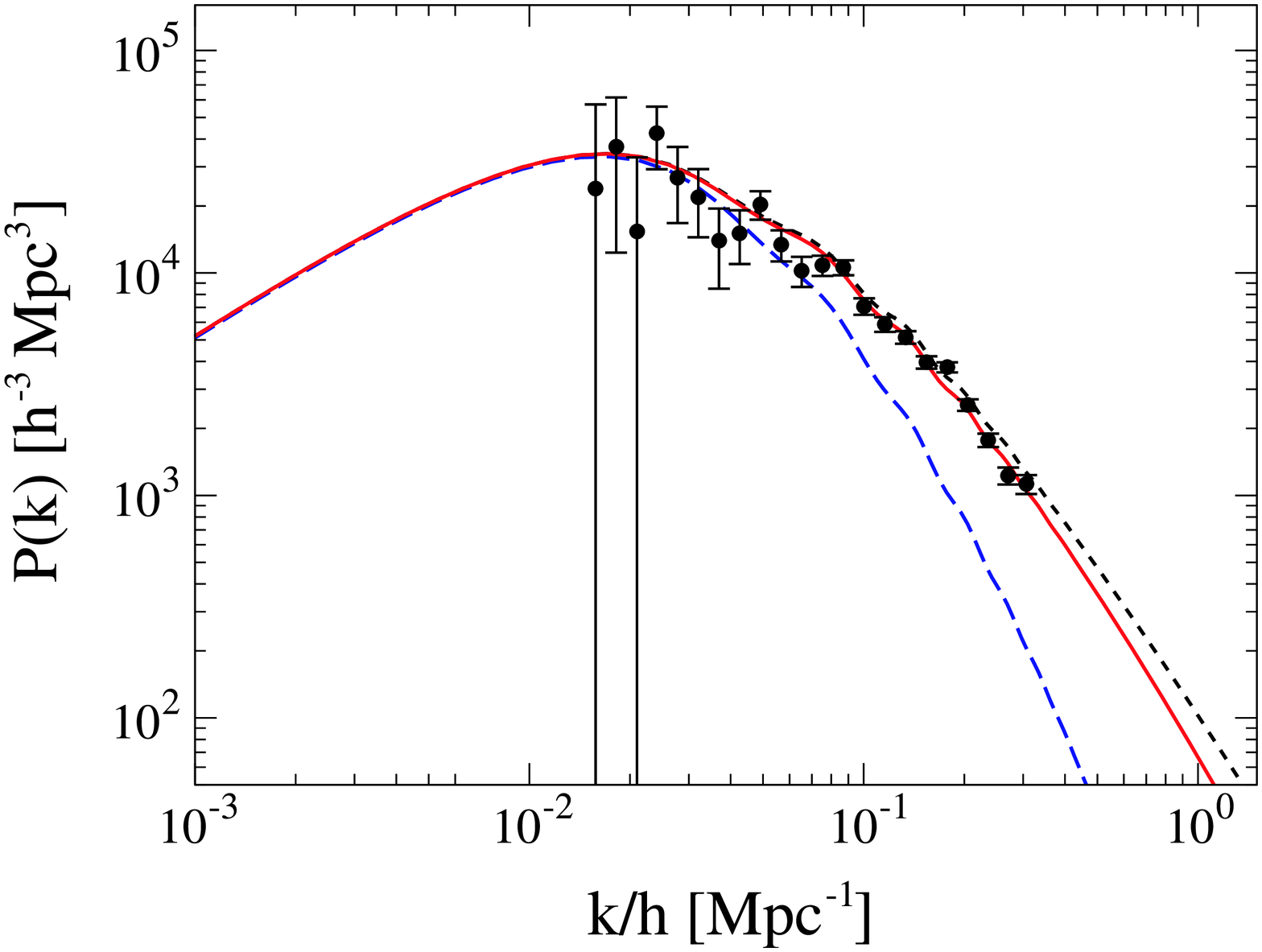}} \\[0.05cm]
\mbox{\bf (a)}\\
\resizebox{7.73cm}{!}{\includegraphics{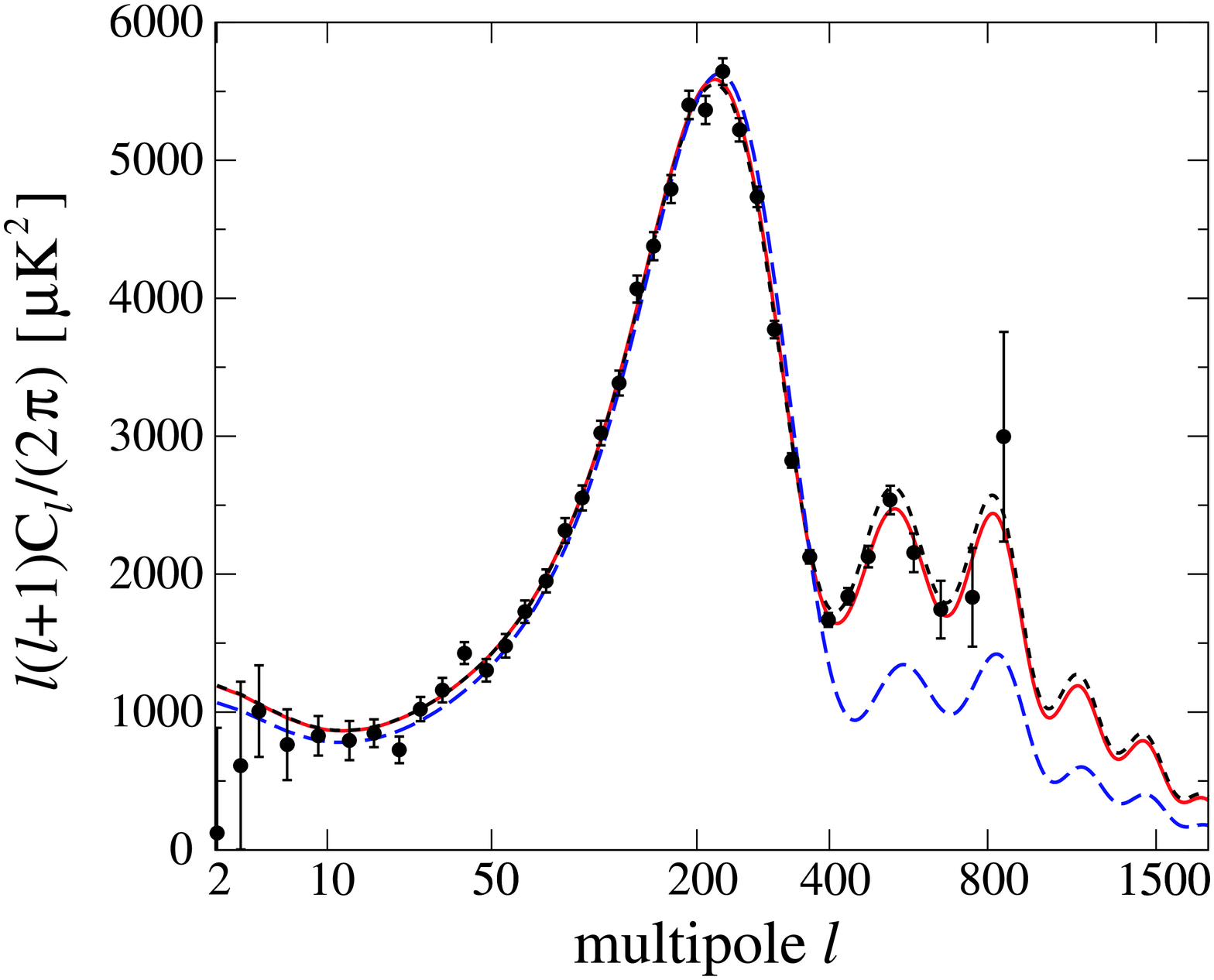}} \\ [0.05cm]
\mbox{\bf (b)}
\end{array}$
\end{center}
\caption{
	The (a) matter (b) CMB power spectra including 
	baryon-$\chi$ drag for $m_{\chi}=1~{\rm GeV}$.   
	The solid curve is for 
$({\cal D}^2+{\cal M}^2)^{1/2}=1.4\times10^{-15}\,e$~cm, 
	short-dashed for 
$({\cal D}^2+{\cal M}^2)^{1/2}=1.0\times10^{-16}\,e$~cm, 
	and long-dashed for 
$({\cal D}^2+{\cal M}^2)^{1/2}=5\times10^{-15}\,e$~cm.  
	The data points are from (a) SDSS 
\cite{Tegmark:2003uf} and (b) WMAP 
\cite{Hinshaw:2003ex}.}
\label{fig:ddmcosmo}
\end{figure}
%%%%%%%%%%%%%%%%%%%%% Fig.5 %%%%%%%%%%%%%%%%%%%%

	In Fig.~\ref{fig:ddmcosmo} we show the linear matter 
	power spectrum and in
 angular power spectrum of the cosmic microwave background (CMB) for several
values of the dipole moment and $m_\chi=1$~GeV.  Using a 
 Markov chain Monte Carlo algorithm (see, e.g.,
Ref.~
\cite{Doran:2003ua}), 
	and data from SDSS 
\cite{Tegmark:2003uf}, 
	WMAP
\cite{Hinshaw:2003ex}, 
	CBI 
\cite{Readhead:2004gy}, 
	VSA 
\cite{Dickinson:2004yr},
	and Type Ia supernovae 
\cite{SnIa} 
	we find the bound shown in Fig. 
\ref{fig:concord}.  
	Dipole moments as large as ${\cal D} \sim 10^{-17}~e~{\rm cm}$ 
	are thus cosmologically viable.

%%%%%%%%%%%%%%%%%%%%% SubSection 2.6 %%%%%%%%%%%%%%%%%%%%%%
\subsection{Gamma Rays}
%%%%%%%%%%%%%%%%%%%%% SubSection 2.6 %%%%%%%%%%%%%%%%%%%%%%

	In the Galactic halo $\chi$--$\bar{\chi}$ pairs can annihilate 
	to two photons through the diagrams shown in Fig. 
\ref{fig:ddannhil}a. 
	The non-observation of a gamma-ray line by EGRET leads 
	to the limit shown in Fig. 
\ref{fig:concord}. 
	A detailed search for a line flux with GLAST may find an observable 
	signature for $m_{\chi} \sim 0.1-1$~GeV and ${\cal D}_{17}\sim10$.
	There may also be constraints to the annihilation rate from the low
        excess heat of Uranus \cite{Mitra:2004fh}.

%%%%%%%%%%%%%%%%%%%%% Section 3 %%%%%%%%%%%%%%%%%%%%%%
\section{The Effect of a Quasistable Charged-NLDP}
\label{sec:chargeddecay}
%%%%%%%%%%%%%%%%%%%%% Section 3 %%%%%%%%%%%%%%%%%%%%%%

	In this section we discuss how the decay of a quasistable 
	charged particle $\phi$ (the NLDP) to a neutral dark-matter 
	particle $\chi$ (the LDP) suppresses the linear power spectrum 
	on scales smaller than the horizon during the decay epoch.  
	Prior to decay, the charged NLDPs couple to and oscillate 
	with the primordial plasma.
	After decay, the plasma is coupled only gravitationally 
	to the LDP.  If all LDPs are produced through the late decay 
	of charged NLDPs, then, as shown in Ref.~
\cite{Sigurdson:2003vy}, 
	the power spectrum is cutoff on small scales.  
	For a lifetime $\tau  \sim 3.5~{\rm yr}$ this would reduce 
	the expected number of dwarf galaxies and may solve 
	the small-scale structure problem of cold-dark-matter theory 
\cite{excessCDMpower}.  
	If, instead,  a fraction $f_{\phi}$ of LDPs are produced through 
	the late decay of charged NLDPs, the linear power spectrum is 
	suppressed only by a factor $(1-f_{\phi})^2$.  
	This suppression might be confused with a negative running 
	of the spectral index $\alpha_{s} \equiv d n_{s}/d{\rm ln}k$ 
	in data that probes the power spectrum on small-scales.  
	We describe these effects in further detail below.

%%%%%%%%%%%%%%%%%%%%% SubSection 3.1 %%%%%%%%%%%%%%%%%%%%%%
\subsection{Charged Decay in the Primordial Plasma}
%%%%%%%%%%%%%%%%%%%%% SubSection 3.1 %%%%%%%%%%%%%%%%%%%%%%

	As $\phi \rightarrow \chi$, the $\phi$/$\chi$ comoving 
	density drops/increases as 
$\rho_{\phi}a^{3}=m_{\phi} n_{\phi_0}e^{-t/\tau}$ 
	and 
$\rho_{\chi}a^{3}=m_{\chi}n_{\chi_0}(1-f_{\phi} e^{-t/\tau})$ 
	respectively.
	Here $n_{\chi_0}=\Omega_{\chi}\rho_{crit}/m_{\chi}$ is the
	comoving density of dark matter, $f_{\phi}$ the fraction 
	produced through $\phi$ decays, $a$ is the scale
	factor, and $t$ is the cosmic time. 
	The charged $\phi$ particles are tightly
	coupled to the baryons through Coulomb scattering.  We can thus 
	describe the combined $\phi$-baryon fluid as a generalized
	baryon-like component $\beta$.

	In the synchronous gauge the perturbation evolution equations 
	are identical to the standard equations for the baryons 
	(see, for example, Ref.~
\cite{Ma:1995ey})  
	with the subscript $b$ replaced by $\beta$. 
For the dark matter $\dot{\delta}_{\chi}  = - i k V_{\chi}-\frac{1}{2}\dot{h} + \lambda_m \frac{\rho_{\phi}}{\rho_{\chi}}\frac{a}{\tau}(\delta_{\beta}-\delta_{\chi})$, and $\dot{V}_{\chi} = -\frac{\dot{a}}{a}V_{\chi} + \lambda_m \frac{\rho_{\phi}}{\rho_{\chi}}\frac{a}{\tau}(V_{\beta}-V_{\chi})$, where $\delta_{\chi}={\delta\rho}_{\chi}/\rho_{\chi}$, $V_{\chi}$ is the bulk velocity, $\lambda_m \equiv m_{\chi}/m_{\phi}$, and an overdot is a derivative with respect to conformal time.
The modifications to photon perturbation evolution are
negligibly small.  

\begin{figure}[ht]
\hspace{-0.25in}
%\begin{center}
%\hspace{-0.5cm}
\resizebox{5.in}{!}{\includegraphics{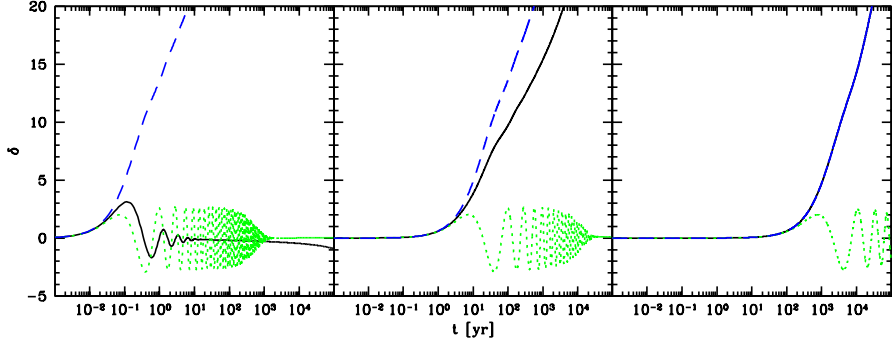}}
\caption{
	The comoving $k=30.0~{\rm Mpc}^{-1}$ (left), 
	$k=3.0~{\rm Mpc}^{-1}$ (center), 
	and $k=0.3~{\rm Mpc}^{-1}$ (right) $\delta_{\chi}$ 
	perturbation for the $\Lambda$CDM model  (long-dashed) 
	and a $f_{\phi}=1$ model with $\tau=3.5~{\rm yr}$ (solid).  
	Also shown is the $\delta_{\beta}$ perturbation (short-dashed).} 
\label{fig:delta_tau}
%\end{center}
\end{figure}

	Due to Compton scattering the $\beta$ component and the photons 
	are tightly coupled as a $\beta$-photon fluid at early times which 
	supports acoustic oscillations.  
	Since dark-matter perturbations are sourced by the
	perturbations of the $\beta$ component, $k$-modes that
	enter the horizon prior to decay, when $\rho_{\phi}/\rho_{\chi}$ 
	is large,  will track the oscillations in the $\beta$-photon 
	fluid rather than growing due to gravity.  
	After decay, when $\rho_{\phi}/\rho_{\chi}$ is small, 
	dark-matter modes that enter the horizon undergo the standard 
	growing evolution. In Fig.~
\ref{fig:delta_tau} 
	we follow the evolution of the dark-matter perturbations through
	the epoch of decay.   We used a modified version of {\tt cmbfast}
\cite{Seljak:1996is}.

%%%%%%%%%%%%%%%%%%%% Fig.6 %%%%%%%%%%%%%%%%%%%%

%%%%%%%%%%%%%%%%%%%%% Fig.6 %%%%%%%%%%%%%%%%%%%%

\begin{figure}[ht]
\begin{center}
$\begin{array}{c@{\hspace{-0.1in}}c}
\resizebox{6.8cm}{!}{\includegraphics{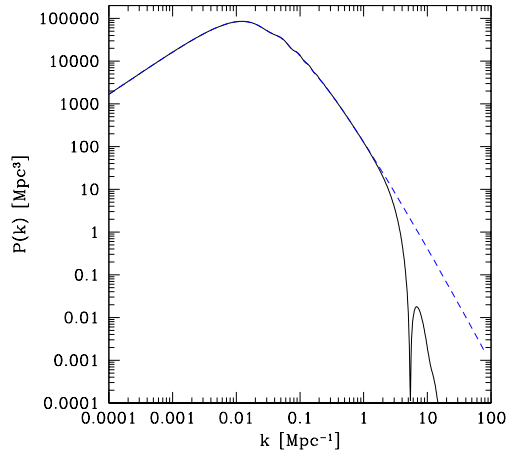}} \\[0.05cm]
\mbox{\bf (a)}\\
\resizebox{6.8cm}{!}{\includegraphics{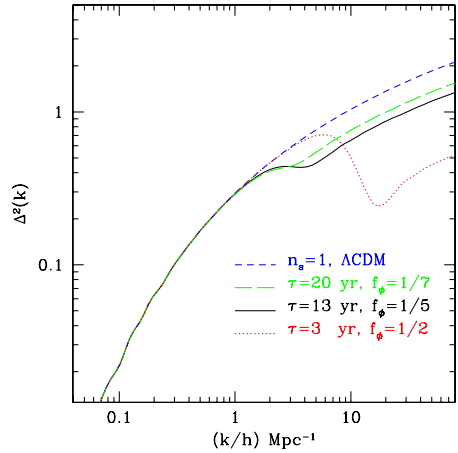}} \\ [0.05cm]
\mbox{\bf (b)}
\end{array}$
\caption{
	(a): The linear power in the standard $\Lambda$CDM model 
	(dashed), and a $f_{\phi}=1$ charged-NLDP model (solid) 
	with $\tau = 3.5~{\rm yr}$.  
	(b): Shown is $\Delta^2(k)=k^3P(k)/2\pi^2$ 
	for a $n_s=1$ $\Lambda$CDM model (dashed) and for several 
	charged-NLDP models. }
\label{fig:pow}
\end{center}
\end{figure}

	In Fig.~\ref{fig:pow}a we plot the
	linear power spectrum for $f_{\phi}=1$ 
	and $\tau  = 3.5~{\rm yr}$. 
	Power is cutoff for $k^{-1} \aplt 0.3~ {\rm Mpc}$ 
	relative to a standard $\Lambda$CDM power
	spectrum, reducing the expected number of
	subgalactic halos and bringing predictions in line with
	observation 
\cite{Kamion2000}.  
	In Fig.~
\ref{fig:pow}b 
	we show the linear power spectrum for several values 
	of $\tau$ and $f_{\phi}$.  The linear power spectrum is now 
	suppressed by a factor of $(1-f_{\phi})^2$.  
	On large scales the linear power spectrum describes 
	the statistics of density fluctuations, but on small scales 
	the full nonlinear matter power spectrum is required.  
	The effects of nonlinear evolution are accounted for in Ref.~
\cite{Profumo:2004qt} 
	where it is shown that the effect of the charged-NLDP decay 
	on the small-scale nonlinear power spectrum can be similar 
	(but different in detail) to that of a model parametrized 
	by a running of the spectral index 
$\alpha_{s} \equiv d n_{s}/d{\rm ln}k$. 

	We note that since the NLDP is charged, LDP production 
	will be accompanied by an electromagnetic cascade.  
	The latter could in principle reprocess the light elements 
	produced during big bang nucleosynthesis, 
	or induce unreasonably large spectral distortions to the CMB.  
	In fact, the models discussed are safely below current limits 
\cite{Sigurdson:2003vy,Profumo:2004qt}.

%%%%%%%%%%%%%%%%%%%%% Fig.7 %%%%%%%%%%%%%%%%%%%%
%%%%%%%%%%%%%%%%%%%%% Fig.7 %%%%%%%%%%%%%%%%%%%%

%%%%%%%%%%%%%%%%%%%%% SubSection 3.1 %%%%%%%%%%%%%%%%%%%
\subsection{Particle-Theory Models of Quasistable Charged NLDPs}
%%%%%%%%%%%%%%%%%%%%% SubSection 3.1 %%%%%%%%%%%%%%%%%%%

	In the $f_{\phi}=1$ case, in order to be a solution 
	to the small-scale structure problem the charged-NLDP 
	must have a comoving density equal to the dark-matter 
	density today and $\tau \sim$ yr.  One way to make the NDLP 
	long lived is to suppress the LDP coupling. 
	If LDP is a stable super-weakly--interacting particle 
\cite{Feng2003}, 
	such as a gravitino $\widetilde{G}$ or the first Kaluza-Klein 
	graviton  $G^1$ in the universal extra-dimensions scenario 
\cite{Servant:2002aq}, 
	the NLDP can be charged and, since it decays gravitationally, 
	have a super-weak decay rate. The required mass-spliting 
	and lifetime can be achieved for 
$m_{\phi} \approx m_{\chi} \sim 100~{\rm TeV}$.  
	These masses are above the ${\rm TeV}$ range discussed 
	in most supersymmetric phenomenology, 
	and uncomfortably close to violating the unitary bound 
	for thermal production 
\cite{Griest:1989wd} --- 
	nonthermal production is likely required for such large masses.

	For $f_{\phi} < 1$, remarkably, there are
	configurations in the minimal supersymmetric extension 
	of the standard model  (MSSM) with the properties required here 
\cite{Profumo:2004qt}. 
	If the LDP is a neutralino quasi-degenerate in mass with 
	the lightest stau, one can naturally 
	obtain, at the same time, LDPs with a relic abundance 
	$\Omega_{\chi}h^2 = 0.113$ 
\cite{CMBconstraints} 
	and NLDP lifetimes and the densities 
	needed in the proposed scenario. Such configurations arise even
	in minimal schemes, such as minimal
	supergravity (mSUGRA)~
\cite{msugra} 
	and the minimal anomaly-mediated supersymmetry-breaking (mAMSB) 
	model~
\cite{mamsb}. 
	A detailed study of the ($\tau$,$f_{\phi}$) parameter space 
	using current and future cosmological data may constrain 
	otherwise viable regions of the MSSM parameter space. 
	Furthermore, this scenario may be tested at future particle 
	colliders (such as the Large Hadron Collider (LHC) ) or dark matter 
	detection experiments (See Ref.~
\cite{Profumo:2004qt} 
	for full details). The decays of staus at rest might even 
	be studied by trapping these particles in large water tanks 
	placed outside of LHC detectors
\cite{Feng:2004yi}.  Other methods of trapping staus have also been considered
\cite{Hamaguchi:2004df}.

%%%%%%%%%%%%%%%%%%%%% Section 4 %%%%%%%%%%%%%%%%%%%
\section{Conclusion}
\label{section7}
%%%%%%%%%%%%%%%%%%%%% Section 4 %%%%%%%%%%%%%%%%%%%

	We have considered two distinct scenarios for electromagnetic 
	interactions in the dark sector.  In the first we considered 
	the effects of a LDP with nonzero electric and/or magnetic 
	dipole moments and found that a light particle with 
	$m_{\chi} \approx$~MeV--GeV and a dipole as large as 
	$\sim 3\times 10^{-16}e$~cm~$\approx 1.6\times10^{-5}\,\mu_B$ 
	satisfies experimental and observational bounds.  
	The second scenario examines how the decay of charged NLDPs 
	in the early Universe results in, depending on $f_{\phi}$, 
	either a cutoff or suppression of the small-scale power spectrum.  
	For a purely gravitationally interacting LDP configurations with 
	$f_{\phi}=1$ may exist, while configurations with $f_{\phi} < 1$ 
	can be found within the MSSM.

%%%%%%%%%%%%%%%%%%%%% Section 5 %%%%%%%%%%%%%%%%%%%
\section{Acknowledgments}
%%%%%%%%%%%%%%%%%%%%% Section 4 %%%%%%%%%%%%%%%%%%%

	We thank Robert R. Caldwell, Michael Doran, Stefano Profumo, Marc 
	Kamionkowski, Andriy Kurylov, and Piero Ullio for collaborative work 
	discussed in this article and the Mitchell Institute and Texas A\&M 
	University for their hospitality during DARK 2004.  
	This research was supported in part by a NSERC 
	of Canada Postgraduate Scholarship, NASA NAG5-9821, 
	and DoE DE-FG03-92-ER40701.
%
%
% BibTeX users please use
% \bibliographystyle{}
% \bibliography{}
%
% Non-BibTeX users please follow the syntax
% the syntax of "referenc.tex" for your own citations
\input{referenc}

%%%%%%%%%%%%%%%%%%%%%%%%%%%%%%%%%%%%%%%%%%%%%%%%%%%%%%%%%%%%%%%%%%%%%%  }

%%%%%%%%%%%%%%%%%%%%%%%%%%%%%%%%%%%%%%%%%%%%%%%%%%%%%%%%%%%%%%%%%%%%%%

\printindex
\end{document}

%% file: referenc.tex
%%%%%%%%%%%%%%%%%%%%%%%% referenc.tex %%%%%%%%%%%%%%%%%%%%%%%%%%%%%%
% sample references
% "physics"
%
% Use this file as a template for your own input.
%
%%%%%%%%%%%%%%%%%%%%%%%% Springer-Verlag %%%%%%%%%%%%%%%%%%%%%%%%%%

%
% BibTeX users please use
% \bibliographystyle{}
% \bibliography{}